\documentclass[preprint,superscriptaddress,preprintnumbers,amsmath,amssymb,prd]{revtex4}
\usepackage{graphicx}

\begin{document}

\title{Impact of magnetic nanoparticles on the Casimir pressure in three-layer systems}

\author{
G.~L.~Klimchitskaya}
\affiliation{Central Astronomical Observatory at Pulkovo of the
Russian Academy of Sciences, Saint Petersburg,
196140, Russia}
\affiliation{Institute of Physics, Nanotechnology and
Telecommunications, Peter the Great Saint Petersburg
Polytechnic University, Saint Petersburg, 195251, Russia}

\author{
V.~M.~Mostepanenko}
\affiliation{Central Astronomical Observatory at Pulkovo of the
Russian Academy of Sciences, Saint Petersburg,
196140, Russia}
\affiliation{Institute of Physics, Nanotechnology and
Telecommunications, Peter the Great Saint Petersburg
Polytechnic University, Saint Petersburg, 195251, Russia}
\affiliation{Kazan Federal University, Kazan, 420008, Russia}

\author{
E. K. Nepomnyashchaya}
\affiliation{Institute of Physics, Nanotechnology and
Telecommunications, Peter the Great Saint Petersburg
Polytechnic University, Saint Petersburg, 195251, Russia}

\author{
E. N. Velichko}
\affiliation{Institute of Physics, Nanotechnology and
Telecommunications, Peter the Great Saint Petersburg
Polytechnic University, Saint Petersburg, 195251, Russia}

\begin{abstract}
The Casimir pressure is investigated in three-layer systems where the intervening
stratum possesses magnetic properties. This subject is gaining in importance in
connection with ferrofluids and their use in various microelectromechanical
devices. We present general formalism of the Lifshitz theory adapted to the case
of ferrofluid sandwiched between two dielectric walls. The Casimir pressure is
computed for the cases of kerosene- and water-based ferrofluids containing a
5$\%$ fraction of magnetite nanoparticles with different diameters between silica
glass walls. For this purpose, we have found the dielectric permittivities of
magnetite and kerosene along the imaginary frequency axis employing the available
optical data and used the familiar dielectric properties of silica glass and water,
as well as the magnetic properties of magnetite. We have also computed the relative
difference in the magnitudes of the Casimir pressure which arises on addition of
magnetite nanoparticles to pure carrier liquids. It is shown that for  nanoparticles
of 20~nm diameter at 2 micrometer separation between the walls this relative
difference exceeds 140$\%$ and 25$\%$ for kerosene- and water-based ferrofluids,
respectively. An interesting effect is found that at a fixed separation between
the walls an addition of magnetite nanoparticles with some definite diameter makes
no impact on the Casimir pressure. The physical explanation for this effect is
provided. Possible applications of the obtained results are discussed.
\end{abstract}

\maketitle

\section{Introduction}
It has long been known that with decreasing distance between two adjacent surfaces
the van der Waals \cite{b1} and Casimir \cite{b2} forces come into play which are
caused by the zero-point and thermal fluctuations of the electromagnetic field.
These forces are of common nature. In fact, the van der Waals force is a special
case of the Casimir force when the separation distance reduces to below a few
nanometers, where the effects of relativistic retardation are negligibly small.
The theory of the van der Waals and Casimir forces was developed by E.~M.~Lifshitz
and his collaborators \cite{b3,b4}. In the framework of this theory, the force
value is expressed via the frequency-dependent dielectric permittivities and
magnetic permeabilities of the boundary surfaces. For ordinary,  nonmagnetic,
surfaces the Casimir force through a vacuum gap is always attractive. If, however,
the gap is filled with a liquid, the Casimir force may be repulsive if the
dielectric permittivities of the boundary surfaces and of a liquid satisfy some
condition \cite{b3,b4}. This is the case of a three-layer system which suggests
a wide variety of different options.

By now many measurements of the Casimir force acting through a vacuum gap have
been performed between both nonmagnetic (see Refs.~\cite{b2,b5,b6} for a review)
and magnetic \cite{b7,b8,b9,b10} materials. The attractive and repulsive forces
in the three-layer systems involving a liquid stratum were measured as well \cite{b11,b12,b13}. The obtained results have been used to devise various
micro- and nanodevices actuated by the Casimir force
\cite{b14,b15,b16,b17,b18,b19,b20,b21,b22,b23,b24}.
All of them, however, exploit the Casimir force through a vacuum gap for
their functionality. At the same time, the so-called magnetic (or ferro)
fluids \cite{b25}, which consist of some carrier liquid and the magnetic
nanoparticles coated with a surfactant to prevent their agglomeration, find
expanding applications in the mechanical engineering, electronic devices, optical
modulators and switchers, optoelectronic communications, biosensors, medical
technologies etc. (see Ref. \cite{b26} for a review and, e.g.,
Refs.~\cite{b27,b28,b29,b30}). Among them, from the viewpoint of Casimir
effect, the applications of greatest interest are to micromechanical sensors
\cite{b31}, microfluidics \cite{b32,b33} and microrobotics \cite{b34}, where
ferrofluids may be confined between two closely spaced surfaces, forming the
three-layer system. Note, however, that the Casimir force in systems of this
kind, having a magnetic intervening stratum, was not investigated so far.

In this paper, we consider an impact of magnetic nanoparticles on the Casimir
pressure in three-layer systems, where the magnetic fluid is confined in between
two glass plates. The cases of kerosene- and water-based ferrofluids are treated
which form a colloidal suspension with magnetite nanoparticles of some diameter
$d$. The Casimir pressure in the three-layer systems with a magnetic intervening
stratum is calculated in the framework of the Lifshitz theory
\cite{b2,b3,b4}.  For this purpose, we find the dielectric permittivity of
kerosene and both the dielectric and magnetic characteristics of magnetite
and ferrofluids at the pure imaginary Matsubara frequencies.

The computational results are presented for both the magnitude of the Casimir
pressure through a ferrofluid and for the impact of magnetic nanoparticles on
the  Casimir pressure through a nonmagnetic fluid. These results are shown as
function of separation distance between the plates and of a nanoparticle
diameter.  The effect of the conductivity of magnetite at low frequencies on
the results obtained is discussed. We show that the presence of magnetic
nanoparticles in the intervening liquid makes a significant impact on the
magnitude of the Casimir pressure. Thus, for the $5\%$ fraction of magnetic
nanoparticles with 20~nm diameter in a kerosene-based ferrofluid, at
$2~\mu$m separation between the walls, this impact exceeds  $140\%$.
For a water-based ferrofluid under the same conditions the presence of
magnetic nanoparticles enhances the magnitude of the Casimir pressure by $25\%$.
Another important result is that at a fixed separation the presence of
magnetic nanoparticles of some definite diameter makes no impact on the
Casimir pressure. The physical reasons for this conclusion are elucidated.

The paper organized as follows. In Sec.~II, we present the formalism of the
Lifschitz theory adapted for a three-layer system with magnetic intervening
stratum.  We also find the dielectric permittivity and magnetic permeability
of magnetite along the imaginary frequency axis and include necessary
information regarding the dielectric permittivity of a colloidal suspension.
Section~III contains evaluation of the dielectric permittivity of kerosene
and kerosene-based ferrofluids along the imaginary frequency axis.
Here we calculate the Casimir pressure in such ferrofluids and investigate
the role of magnetite nanoparticles in the obtained results.  In Sec.~IV
the same is done for the case of water-based ferrofluids.  In Sec.~V the
reader will find our conclusions and a discussion.

\section{General formalism for three-layer systems with
{\protect{\\}} magnetite nanoparticles}

We consider the three-layer system consisting of two parallel nonmagnetic
dielectric walls described by the frequency-dependent dielectric permittivity
$\varepsilon(\omega)$ and separated by the gap of width $a$. The gap is filled
with a ferrofluid having the dielectric permittivity $\varepsilon_{\rm ff}(\omega)$
and magnetic permeability $\mu_{\rm ff}(\omega)$. The thickness of the walls is taken
to be sufficiently large in order they could be considered as semispaces.
This is the case for the dielectric walls with more than 2~$\mu$m thickness
\cite{b35}. Then, assuming that our system is in thermal equilibrium with the
environment at temperature $T$, the Casimir pressure between the walls can be
calculated by the Lifshitz formula \cite{b2,b3,b4}
\begin{eqnarray}
&&
{P}(a)=-\frac{k_BT}{\pi}\sum_{l=0}^{\infty}
\vphantom{\sum}^{'}\int_0^{\infty}k_{\bot}dk_{\bot}k_{\rm ff}(i{\xi}_l, k_{\bot})
\label{eq1}\\
&&~~~
\times \sum_{\alpha}\left[\frac{e^{2{\alpha}k_{\rm ff}(i{\xi}_l, k_{\bot})}}{r^2_{\alpha}(i{\xi}_l, k_{\bot})}-1\right]^{-1}.
\nonumber
\end{eqnarray}
\noindent
Here, $k_B$ is the Boltzmann constant, $\xi_l=2{\pi}k_BTl/\hbar$, where
$l=0,\,1,\,2,\,\ldots\,$, are the Matsubara frequencies, the prime on the
summation sign in $l$ divides the term with $l=0$ by $2$, $k_{\bot}$ is
the magnitude of the wave vector projection on the plane of walls, and
\begin{equation}\label{eq2}
  k_{\rm ff}(i\xi_l,k_{\bot})=\left[k_{\bot}^2+
  \varepsilon_{\rm ff}(i\xi_l)\mu_{\rm ff}(i\xi_l)\frac{\xi_l^2}{c^2}\right]^{1/2}.
\end{equation}
\noindent
The reflection coefficients $r_\alpha{(i\xi_l,k_{\bot})}$ in our three-layer
system are defined for two independent polarizations of the electromagnetic
field, transverse magnetic ($\alpha={\rm TM}$) and transverse electric
($\alpha={\rm TE}$). They are given by \cite{b2}
\begin{eqnarray}
&&
r_{\rm TM}(i\xi_l,k_{\bot})=\frac{\varepsilon(i\xi_l)k_{\rm ff}(i\xi_l,k_{\bot})-
\varepsilon_{\rm ff}(i\xi_l)k(i\xi_l,k_{\bot})}{\varepsilon(i\xi_l)k_{\rm ff}(i\xi_l,k_{\bot})+
\varepsilon_{\rm ff}(i\xi_l)k(i\xi_l,k_{\bot})},
\nonumber\\
&&
r_{\rm TE}(i\xi_l,k_{\bot})=\frac{k_{\rm ff}(i\xi_l,k_{\bot})-
\mu_{\rm ff}(i\xi_l)k(i\xi_l,k_{\bot})}{k_{\rm ff}(i\xi_l,k_{\bot})+
\mu_{\rm ff}(i\xi_l)k(i\xi_l,k_{\bot})},
\label{eq3}
\end{eqnarray}
\noindent
where we have introduced the standard notation
\begin{equation}
k(i\xi_l,k_{\bot})=\left[k_{\bot}^2+\varepsilon(i\xi_l)\frac{\xi_l^2}{c^2}\right]^{1/2}.
\label{eq4}
\end{equation}

\begin{figure}[b]
\vspace*{-4cm}
\centerline{\hspace*{2.5cm}
\includegraphics{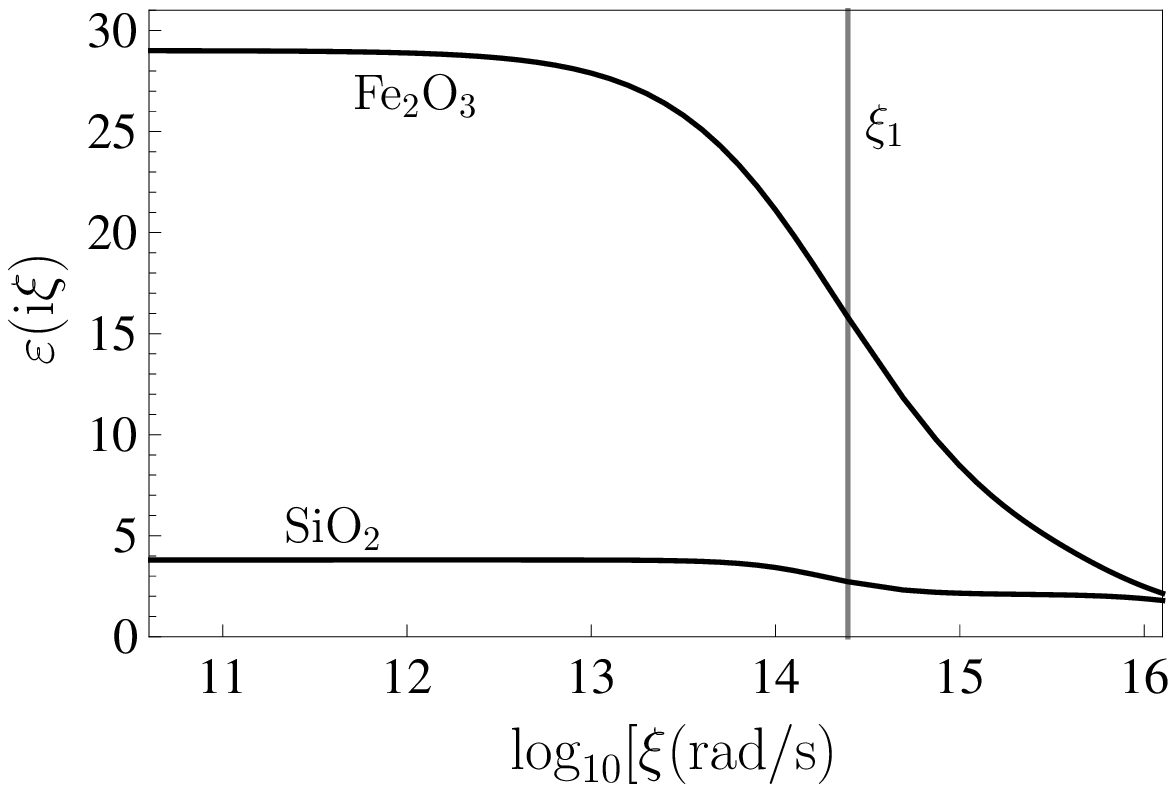}
}
\vspace*{-9.5cm}
\caption{\label{fg1}
The dielectric permittivities of magnetite nanoparticles and silica glass are shown as the functions of imaginary frequency by the top and bottom lines, respectively.  The vertical line indicates the position of the first Matsubara frequency.
}
\end{figure}

As is seen from Eqs.~(\ref{eq1})--(\ref{eq4}),
calculation of the Casimir pressure in the three-layer system is straightforward
if one knows the dielectric and magnetic properties of all layers described by
the functions $\varepsilon(i\xi_l)$, $\varepsilon_{\rm ff}(i\xi_l)$ and
$\mu_{\rm ff}(i\xi_l)$. The dielectric permittivity $\varepsilon(i\xi)$ of a silica
glass, considered in the next sections as the material of walls, has been
much studied \cite{b2,b36}. It is shown by the bottom line in Fig.~\ref{fg1} as the
function of $\xi$.  Specifically, at zero frequency one has $\varepsilon(0)=3.801$.
The ferrofluid is a binary mixture of nanoparticles plus a carrier liquid.
Here, we consider the dielectric and magnetic properties of magnetite
Fe$_3$O$_4$ nanoparticles which make an intervening liquid stratum magnetic.

The real and imaginary parts of the dielectric permittivity of
magnetite $\varepsilon_m(\omega)$ have been measured in a Ref.~\cite{b37}
in the frequency region from $\Omega_1=2\times10^{14}$~rad/s to
$\Omega_2=1.8\times10^{16}$~rad/s (i.e., from $\hbar\Omega_1=0.13$~eV to $\hbar\Omega_2=12$~eV).
We have extrapolated the measurement results of Ref.~\cite{b37} for
${\rm Im}\varepsilon_m(\omega)$ to the region of lower frequencies $\omega<\Omega_1$ by using the imaginary part of the Debye permittivity
\begin{equation}\label{eq5}
{\rm Im} \varepsilon_{m}(\omega)=\frac{C_D\,\omega_D\,\omega}{\omega_D^2+\omega^2}.
\end{equation}
\noindent
The values of two parameters $C_D=24.02$ and $\omega_D=2.05\times10^{14}$~rad/s
were determined from the condition of smooth joining between the measured data
and the Debye extrapolation.  An extrapolation to the region of higher frequencies $\omega>\Omega_2$ was done by means of the standard theoretical dependence
\begin{equation}\label{eq6}
{\rm Im} \varepsilon_{m}(\omega)=C\left(\frac{\Omega_2}{\omega}\right)^3,
\end{equation}
\noindent
where the experimental data at high frequencies lead to
$C= 1.58$.

Now we substitute Eqs.~(\ref{eq5}) and (\ref{eq6}) in the right-hand side of
the  Kramers-Kronig relation \cite{b2} and obtain
\begin{equation}
\varepsilon_m(i\xi)=1+\frac{2}{\pi}\left[I_1(\xi)+I_2(\xi)+I_3(\xi)\right],
\label{eq7}
\end{equation}
\noindent
where
\begin{eqnarray}
&&
I_1(\xi)=C_D\omega_D\int_0^{\Omega_1}
\frac{\omega^2d\omega}{(\omega_D^2+\omega^2)(\xi^2+\omega^2)},
\nonumber\\[1mm]
&&
I_2(\xi)=\int_{\Omega_1}^{\Omega_2}
\frac{\omega{\rm Im}\varepsilon_m(\omega)}{\xi^2+\omega^2}d\omega,
\label{eq8}\\[1mm]
&&
I_3(\xi)=C\Omega_2^3\int_{\Omega_2}^{\infty}
\frac{d\omega}{\omega^2(\xi^2+\omega^2)}
\nonumber
\end{eqnarray}
\noindent
and ${\rm Im}\varepsilon_m(\omega)$ in $I_2(\xi)$ is given by the measurement data of Ref.~\cite{b37}.

Calculating the integrals $I_1(\xi)$ and $I_3(\xi)$, one finds
\begin{eqnarray}
&&
I_1(\xi)=\frac{C_D\omega_D^2}{\xi^2-\omega_D^2}
\left[\frac{\xi}{\omega_D}\text{arctan}\frac{\Omega_1}{\xi}-
\text{arctan}\frac{\Omega_1}{\omega_D}\right],
\nonumber\\
&&
I_3(\xi)=\frac{C\Omega_{2}^{2}}{\xi^2}
\left[1+\frac{\Omega_2}{\xi}\text{arctan}\frac{\Omega_2}{\xi}-\frac{\pi}{2}\right].
\label{eq9}
\end{eqnarray}

Note that in the limiting case $\xi/\Omega_2\ll1$ one has
\begin{equation}
  I_3(\xi)=\frac{1}{3}C\left(1-\frac{3}{5}\frac{\xi^2}{\Omega_2^2}+
  \frac{3}{7}\frac{\xi^4}{\Omega_2^4}\right),
\label{eq10}
\end{equation}
\noindent
and, thus, $I_3(0)=C/3\approx0.53$. This is much smaller than $I_1(0)=C_D\text{arctan}(\Omega_1/\omega_D)\approx18.38$ and, as it follows
from numerical computations, than $I_2(0)\approx25.07$. In such a manner the
region of high real frequencies gives only a minor contribution to $\varepsilon_m(0)=29.0$.

Using Eqs.~(\ref{eq7}) and (\ref{eq8}) we have calculated $\varepsilon_m$ as
a function of $\xi$.  The computational results are shown by the top line
in Fig.~\ref{fg1}.  In the same figure, the position of the first Matsubara
frequency $\xi_1$ is indicated by the vertical line.  Note that in the region
of very low frequencies $\omega\lesssim10^3$~Hz the dielectric permittivity
of magnetite increases significantly together with its electric conductivity
\cite{b38}.  An increase of ${\rm Im}\varepsilon_m(\omega)$ at so low frequencies
does not influence on the values of $\varepsilon_m(i\xi)$ in the frequency
region shown in Fig.~\ref{fg1} and, hence, on the values of
$\varepsilon_m(i\xi_l)$ with $l\geqslant1$.  The conductivity of magnetite
at low frequencies makes an impact only on the term of Eq.~(\ref{eq1}) with
$l=0$ leading to $\varepsilon_m(\xi)\rightarrow\infty$ when $\xi\rightarrow0$.
Below in Secs.~III and IV we consider both options $\varepsilon_m(0)=29.0$
and $\varepsilon_m(0)=\infty$ and adduce the arguments why the former option is
more realistic in computation of the Casimir pressure.

To obtain the dielectric permittivity of the ferrofluid, $\varepsilon_{\rm ff}$,
one should combine the dielectric permittivity of a carrier liquid, $\varepsilon_c$,
with the dielectric permittivity of magnetic nanoparticles, $\varepsilon_m$,
taking into account the volume fraction of the latter $\Phi$ in the ferrofluid.
The permittivity $\varepsilon_c$ is discussed in Secs.~III and IV for different
carrier liquids.  As to the combination law, for the case of spherical
nanoparticles
it is given by the Rayleigh mixing formula \cite{b39} used here for $\omega=i\xi$
\begin{equation}
  \frac{\varepsilon_{\rm ff}{(i\xi)}-
\varepsilon_{c}{(i\xi)}}{\varepsilon_{\rm ff}{(i\xi)}+
2\varepsilon_{c}{(i\xi)}}=
\Phi\frac{\varepsilon_{m}{(i\xi)}-
\varepsilon_{c}{(i\xi)}}{\varepsilon_{m}{(i\xi)}+
2\varepsilon_{c}{(i\xi)}}.
\label{eq11}
\end{equation}
\noindent
Note that Eq.~(\ref{eq11}) is derived under a condition that the nanoparticles
diameter is $d\ll\lambda$,  where $\lambda$ is the characteristic wavelength.
In the region of separations $a\geqslant200~$nm considered below the
contributing frequencies are $\xi\lesssim10^{16}~$rad/s which correspond to
the wavelengths $\lambda\gtrsim180~$nm.  Thus, for nanoparticles with
$d<20$~nm diameter the above condition is largely satisfied.
As mentioned in Sec.~I, magnetic nanoparticles may be coated with some
surfactant to prevent their agglomeration.
Below we assume that the dielectric function of a surfactant is close to that
of a carrier liquid so that ferrofluid can be considered as a mixture of two
substances.

Now we consider the magnetic permeability of a ferrofluid $\mu_{\rm ff}(i\xi_l)$.
First of all it should be stressed that the magnetic properties influence the
Casimir force only through the zero-frequency term of the Lifshitz formula
\cite{b40}.  This is because at room temperature the magnetic permeability
turns into unity at much smaller frequencies than the first Matsubara frequency.
Thus, the quantity of our interest is
\begin{equation}\label{eq12}
  \mu_{\rm ff}(0)=1+4\pi\chi_{\rm ff}(0),
\end{equation}
\noindent
 where the initial susceptibility of a paramagnetic (superparamagnetic) system is given by \cite{b41}
\begin{equation}\label{eq13}
   \chi_{\rm ff}(0)=N\frac{M^2}{3k_BT},
 \end{equation}
\noindent
 where $N=\Phi/V$, $V=\pi d^3/6$ is the volume of a single-domain nanoparticle,
 $M=M_SV$ is its magnetic moment, and $M_S$ is the saturation magnetization per
 unit volume.

It was found that for nanoparticles $M_S$ takes a smaller value than for a bulk
material. Thus, for a bulk magnetite
$M_S\approx~460~\mbox{emu/cm}^3=4.6\times 10^5~$A/m \cite{b42},
whereas for a single magnetite nanoparticle
$M_S\approx300~\mbox{emu/cm}^3=3\times 10^5~$A/m
\cite{b43}.
Substituting Eq.~(\ref{eq13}) in Eq.~(\ref{eq12}), we arrive at
\begin{equation}\label{eq14}
  \mu_{\rm ff}(0)=1+\frac{2\pi^2\Phi}{9}\frac{M_S^2d^3}{k_BT}.
\end{equation}
\noindent
{}From this equation with the volume fraction of nanoparticles $\Phi=0.05$
one finds $\mu_{\rm ff}(0)\approx1.24$ and $2.9$ for magnetite nanoparticles with
$d=10$ and $20$~nm diameter, respectively.  These results do not depend on
the type of carrier liquid.

\section{Impact of magnetite nanoparticles on the Casimir {\protect{\\}} pressure in kerosene-based interlayer}

Kerosene is often used as a carrier liquid in ferrofluids \cite{b44,b45}.  By now
the dielectric properties of kerosene are not sufficiently investigated.
We have applied the measurement data for the imaginary part of the
dielectric permittivity of kerosene in the microwave \cite{b44} and infrared
\cite{b46} regions and the Kramers-Kronig relation to obtain the
Ninham-Parsegian representation for this dielectric permittivity along
the imaginary frequency axis
\begin{equation}\label{eq15}
  \varepsilon_c(i\xi)=1+\frac{B}{1+\xi\tau}+
  \frac{C_{\rm IR}}{1+(\frac{\xi}{\omega_{\rm IR}})^2}+
  \frac{C_{\rm UV}}{1+(\frac{\xi}{\omega_{\rm UV}})^2}.
\end{equation}
\noindent
Here, the second term on the right-hand side describes the contribution to the
dielectric permittivity from the orientation of permanent dipoles in polar liquids.
The value of $B=0.020$ and $1/\tau=8.0\times10^8$~rad/s  were determined from
the measurement data of Ref.~\cite{b44} in the microwave region.  The third term
on the right-hand side of
Eq.~(\ref{eq15}) describes the effect of ionic polarization.  The respective
constants $C_{\rm IR}=0.007$ and $\omega_{\rm IR}=2.14\times10^{14}$~rad/s were
found using the infrared optical data \cite{b46}.  Taking into account that for
kerosene the optical data in the ultraviolet region are missing, the parameters
 of the last, fourth, term on the right-hand side of Eq.~(\ref{eq15}) have been determined following the approach of
Ref.~\cite{b36} with regard to the known value of the dielectric permittivity
of zero-frequency $\varepsilon_c(0)=1.8$ \cite{b44}.
As a result, the values of $C_{\rm UV}=0.773$ and
$\omega_{\rm UV}=1.0\times10^{16}$~rad/s were obtained.

\begin{figure}[b]
\vspace*{-4cm}
\centerline{\hspace*{2.5cm}
\includegraphics{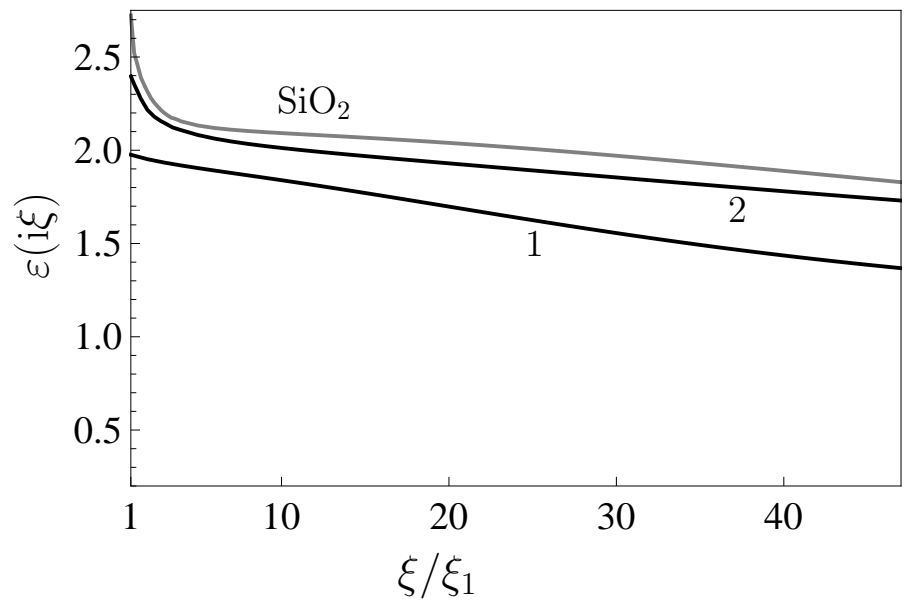}
}
\vspace*{-9.5cm}
\caption{\label{fg2}
The dielectric permittivities of the kerosene- and water-based ferrofluids with 5$\%$ concentration of magnetite nanoparticles (the lines 1 and 2, respectively) and of silica glass are shown as the functions of imaginary frequency normalized to the first Matsubara frequency.
}
\end{figure}

Now the dielectric permittivity $\varepsilon_{\rm ff}(i\xi)$ of kerosene-based
ferrofluid with $\Phi=0.05~(5\%)$  volume fraction of magnetite nanoparticles
is obtained from Eq.~(\ref{eq11}) by substituting the data of the top line
in Fig.~\ref{fg1} for the dielectric permittivity of magnetite
$\varepsilon_m(i\xi)$ and the dielectric permittivity of kerosene
 $\varepsilon_c(i\xi)$ from Eq.~(\ref{eq15}).
The permittivity $\varepsilon_{\rm ff}$ is shown as a function of $\xi/\xi_1$
by the line labeled 1 in Fig.~\ref{fg2}.  In the same figure the dielectric
permittivity of SiO$_2$ walls is reproduced by the top line from Fig.~\ref{fg1}
as a function of $\xi/\xi_1$ in the frequency region important for computations
of the Casimir pressure (the line labeled 2 in Fig.~\ref{fg2} is discussed
in Sec.~IV).  The static dielectric permittivity of a ferrofluid, which is
not shown in the scale of Fig.~\ref{fg2}, is equal to
$\varepsilon_{\rm ff}(0)=2.035$ if the conductivity of magnetite nanoparticles
is disregarded and $\widetilde{\varepsilon}_{\rm ff}(0)=2.084$ if this conductivity
is included in calculations.

Numerical computations of the Casimir pressure are performed most conveniently
by using the dimensionless variables
\begin{equation}\label{eq16}
  y=2ak_{\rm ff}(i\xi_l,k_{\bot}),\quad
  \zeta_l=\frac{\xi_l}{\omega_{\rm cr}}\equiv\frac{2a\xi_l}{c}.
\end{equation}
\noindent
In terms of these variables Eq.~(\ref{eq1}) takes the form
\begin{eqnarray}
&&
  {P}(a)=-\frac{k_BT}{8{\pi}a^3}\sum_{l=0}^{\infty}
\vphantom{\sum}^{'}\int_{\sqrt{\varepsilon_{{\rm ff},l}\mu_{{\rm ff},l}}\zeta_l}^{\infty}
y^2dy
\label{eq17}\\
&&~~~
  \times \sum_{\alpha}\left[\frac{e^y}{r^2_{\alpha}(i{\zeta}_l, y)}-1\right]^{-1},
  \nonumber
\end{eqnarray}
\noindent
where $\varepsilon_{{\rm ff},l}=\varepsilon_{\rm ff}(i\omega_{\rm cr}\zeta_l)$,
 $\mu_{{\rm ff},l}=\mu_{\rm ff}(i\omega_{\rm cr}\zeta_l)$ and the reflection
 coefficients
(\ref{eq3}) are given by
\begin{eqnarray}
&&
r_{\rm TM}(i\zeta_l,y)=\frac{\varepsilon_ly-\varepsilon_{{\rm ff},l}\sqrt{y^2+
(\varepsilon_l-\varepsilon_{{\rm ff},l}\mu_{{\rm ff},l})\zeta_l^2}}{\varepsilon_ly+
\varepsilon_{{\rm ff},l}\sqrt{y^2+(\varepsilon_l-\varepsilon_{{\rm ff},l}\mu_{{\rm ff},l})\zeta_l^2}},
\nonumber\\
&&
r_{\rm TE}(i\zeta_l,y)=\frac{y-\mu_{{\rm ff},l}\sqrt{y^2+(\varepsilon_l-
\varepsilon_{{\rm ff},l}\mu_{{\rm ff},l})\zeta_l^2}}{y+\mu_{{\rm ff},l}\sqrt{y^2+
(\varepsilon_l-\varepsilon_{{\rm ff},l}\mu_{{\rm ff},l})\zeta_l^2}}.
\label{eq18}
\end{eqnarray}
\noindent
with a similar notation for $\varepsilon_l=\varepsilon(i\omega_{\rm cr}\zeta_l)$.

\begin{figure}[b]
\vspace*{-4cm}
\centerline{\hspace*{2.5cm}
\includegraphics{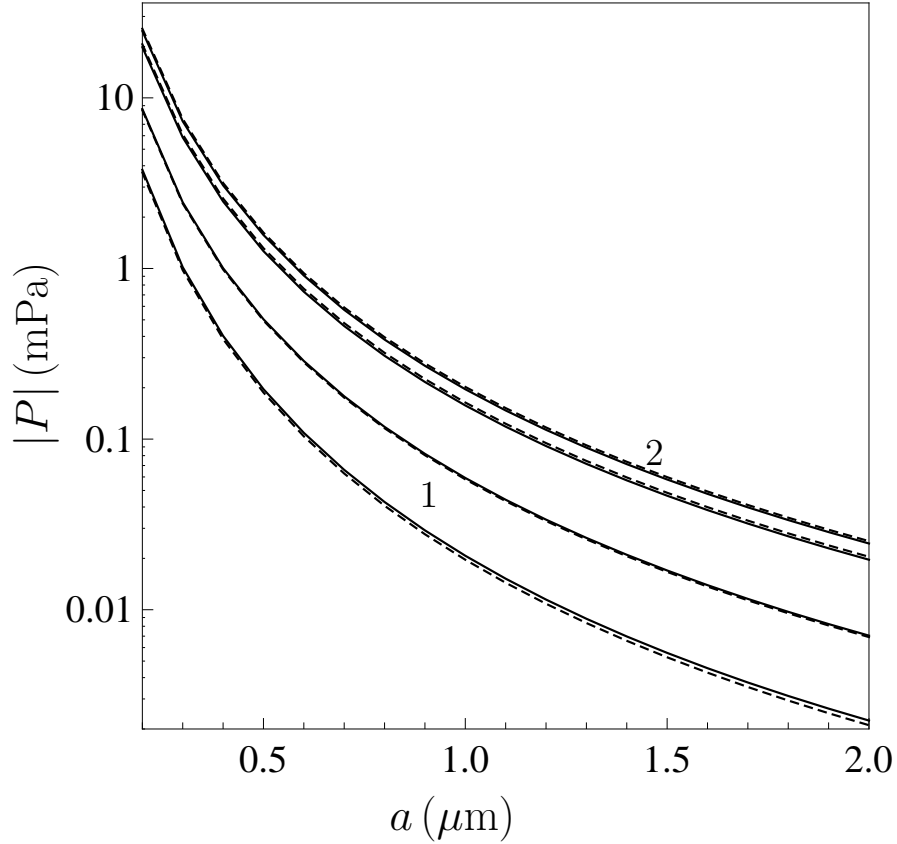}
}
\vspace*{-9.5cm}
\caption{\label{fg3}
The magnitudes or the Casimir pressure between SiO$_2$ walls through a ferrofluid
with 5$\%$ fraction of magnetite nanoparticles are shown as the functions of
separation between the walls by the pairs of solid and dashed lines labeled 1
and 2 for the kerosene and water carrier liquids, respectively.  Solid and dashed
lines are computed with disregarded and included conductivity of magnetite at
low frequencies, respectively.  In each pair the lower line is for nanoparticles
with $d=10$~nm diameter and the upper line
is for nanoparticles with $d=20$~nm.
}
\end{figure}

The magnitude of the Casimir pressure between two SiO$_2$ walls through
the kerosene-based ferrofluid was computed by using Eqs.~(\ref{eq17}) and
(\ref{eq18})
where the dielectric permittivities of SiO$_2$ and of a ferrofluid are given by
the top line and line~1 in Fig.~\ref{fg2}, respectively.
The magnetic permeability of a ferrofluid obtained from Eq.~(\ref{eq14}) in the
end of Sec.~II has been used.  The computational results are shown in Fig.~\ref{fg3}
as functions of separation by the pair of solid lines labeled 1 where the lower and
upper lines are for magnetite nanoparticles with $d=10$ and $20$~nm diameter, respectively.
(The pair of lines labeled 2 is discussed in Sec.~IV).  Note that the
dielectric permittivities used in computations do not depend on the nanoparticle
diameter $d$ which influences the computational results exclusively through the
static magnetic
permeability of the ferrofluid.

The solid lines in pair 1 are calculated with disregarded conductivity of magnetite
at low frequencies, i.e., by using the static dielectric permittivity of a
ferrofluid $\varepsilon_{\rm ff}(0)=2.035$.  The point is that the theoretical
results obtained with taken into account conductivity of dielectric materials
at low frequencies have been found in serious disagreement with the experimental
data of several measurements of the Casimir force
\cite{b2,b5,b47,b48,b49,b50,b51,b52}.  Moreover, the Casimir entropy calculated
with included conductivity at low frequencies was demonstrated to violate the
third law of thermodynamics, the Nernst heat theorem, by taking nonzero positive
value depending on the parameters of a system at zero temperature
\cite{b53,b54,b55,b56}.  Since the deep physical reasons for this experimental
and theoretical conundrum remain unknown, here the computations of the Casimir
pressure are also performed with taken into account conductivity of magnetite
at low frequencies.

The respective computational results are shown in Fig.~\ref{fg3} by the pair of
dashed lines labeled 1.  The lower and upper dashed lines are computed using the
same expressions, as the solid lines, but by using the static dielectric permittivity
of magnetite $\tilde\varepsilon_{\rm ff}(0)=2.084$ for nanoparticles with $d=10$ and $20$~nm, respectively.  As is seen in Fig.~\ref{fg3}, an account for the conductivity of
magnetite at low frequencies makes only a minor impact on the Casimir pressure
in the three-layer system. Thus, for nanoparticles with $d=10$~nm  diameter the
pressures computed at $a=200$~nm with disregarded and included conductivity at
low frequencies are equal to $P=-3.789$~mPa and $\tilde{P}=-3.653$~mPa.
At $a=2~\mu$m similar results are given by $P=-2.24\times10^{-3}$~mPa
and $\widetilde{P}=-2.104\times10^{-3}$~mPa.
For magnetite nanoparticles with $d=20$~nm diameter the Casimir pressures
computed with disregarded and included conductivity at low frequencies
are
$P=-8.598$~mPa and $\tilde{P}=-8.462$~mPa at $a=200$~nm and
$P=-7.049\times10^{-3}$~mPa and $\tilde{P}=-6.913\times10^{-3}$~mPa
at $a=2~\mu$m.

\begin{figure}[b]
\vspace*{-4cm}
\centerline{\hspace*{2.5cm}
\includegraphics{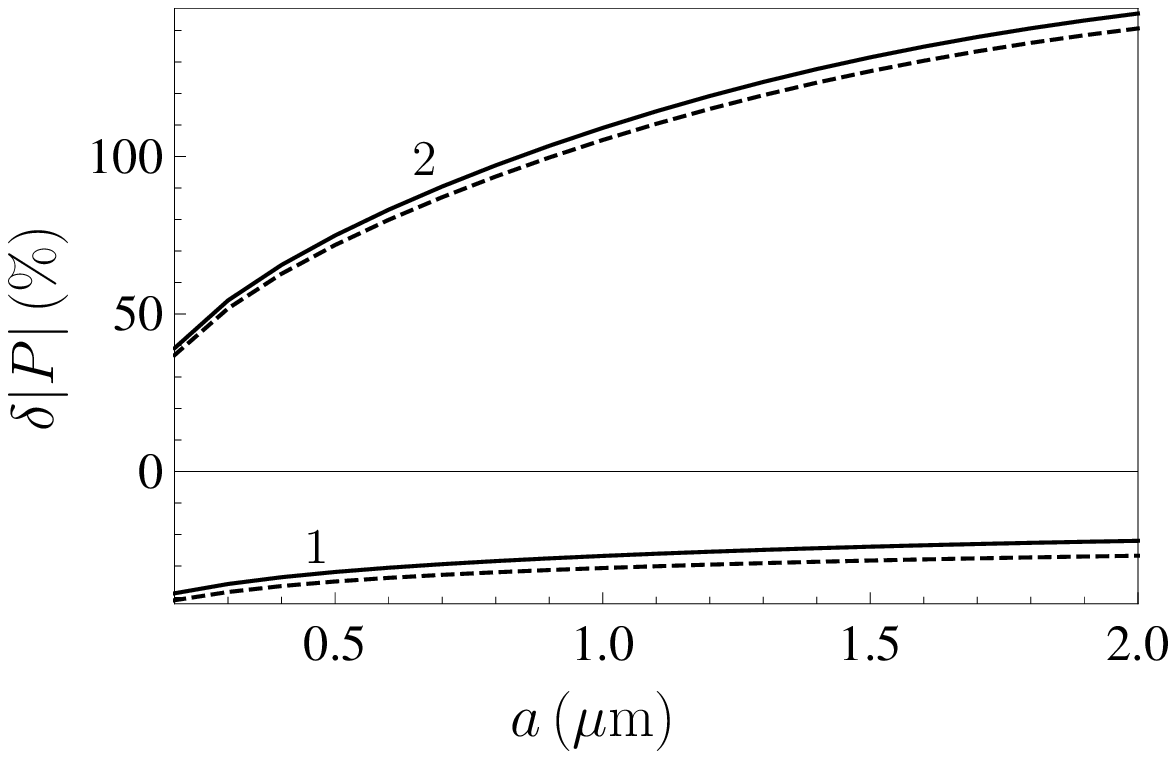}
}
\vspace*{-9.5cm}
\caption{\label{fg4}
 The relative change in the magnitude of the Casimir pressure on addition of the
 5$\%$ fraction of magnetite nanoparticles to kerosene is shown as the function
 of separation by the pairs of solid and dashed lines labeled 1 and 2 for
 nanoparticles with $d=10$~nm and 20~nm diameter, respectively.  In each pair,
 the solid and dashed lines are computed with disregarded and included conductivity
 of magnetite at low frequencies, respectively.
}
\end{figure}

It is interesting to determine the relative impact of magnetite nanoparticles
on the magnitude of the Casimir pressure in a three-layer system.  For this
purpose we have computed the quantity
\begin{equation}\label{eq19}
  \delta|P|=\frac{|P|-|P_{\rm ker}|}{|P_{\rm ker}|},
\end{equation}
\noindent
where $|P_{\rm ker}|$ is the magnitude of the Casimir pressure between two
SiO$_2$
walls through a pure kerosene stratum with no nanoparticles.  The computational
results for the quantity $ \delta|P|$ are shown in Fig.~\ref{fg4} as the functions
of separation
by the pairs of line labeled 1 and 2 for nanoparticles of $d=10$ and $20$~nm
diameter, respectively.  The solid and dashed lines in each pair are computed
with disregarded and included conductivity of magnetite at low frequencies,
respectively.  As is seen in Fig.~\ref{fg4}, on addition of magnetite nanoparticles
with $d=10$~nm diameter to kerosene, the magnitude of the Casimir pressure
decreases.  However, on addition to kerosene of nanoparticles with by a factor of
two larger diameter, the magnitude of the Casimir pressure increases.  Specifically,
at $a=200$~nm one obtains $ \delta|P|=-38.7\%$ and $40.05\%$ for nanoparticles
with $d=10$ and $20$~nm if the conductivity at low frequencies is disregarded
in computations.  If this conductivity is taken into account, the respective results
are $ \delta|\tilde{P}|=-40.9\%$ and $37.8\%$.  The relative impact of
magnetic nanoparticles on the Casimir pressure essentially depends on the
separation between the plates.  Thus, at $a=2~\mu$m $ \delta|P|=-22.0\%$
and $147\%$
for nanoparticles with $d=10$ and $20$~nm if the conductivity of magnetite is
disregarded and $ \delta|\widetilde{P}|=-26.8\%$ and $142\%$ for the same
respective diameters if the conductivity is included in computations.

\begin{figure}[b]
\vspace*{-1cm}
\centerline{\hspace*{2.5cm}
\includegraphics{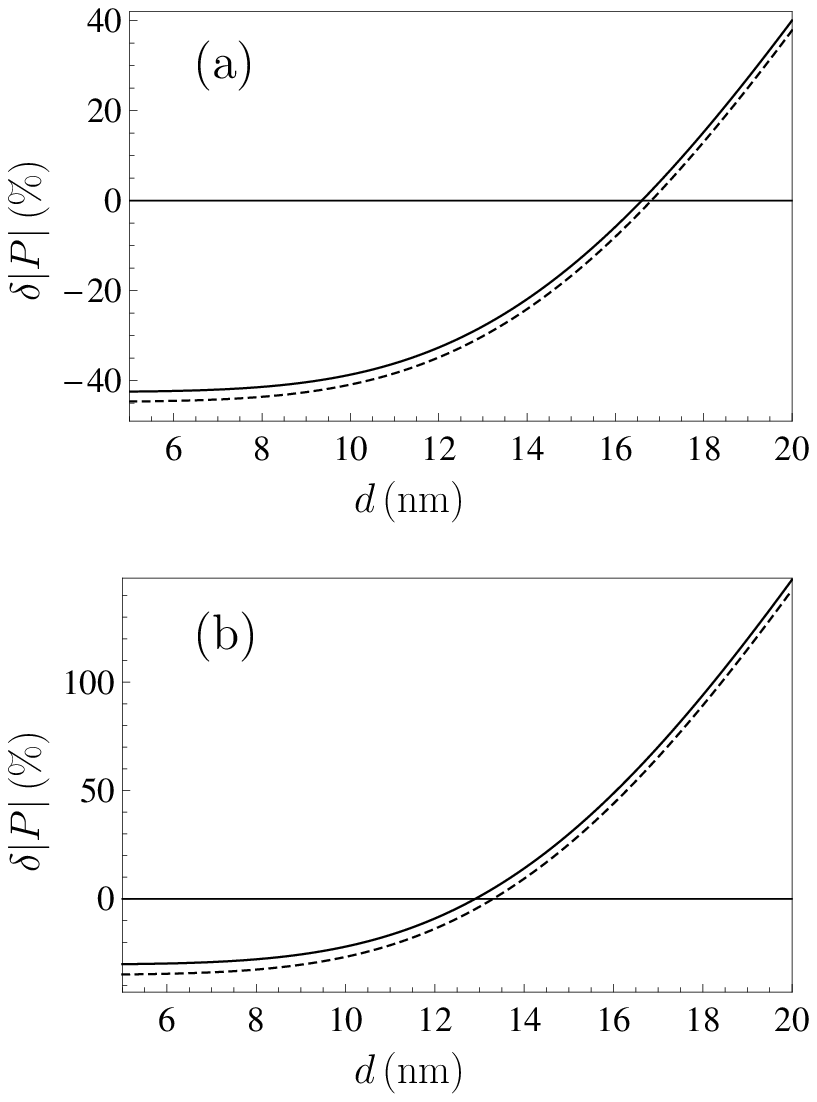}
}
\vspace*{-13.5cm}
\caption{\label{fg5}
The relative change in the magnitude of the Casimir pressure on addition of the
5$\%$ fraction of magnetite nanoparticles to kerosene is shown as the function
of nanoparticle diameter by the pairs of solid and dashed lines computed with
disregarded and included conductivity of magnetite at low frequencies,
respectively, for separation between SiO$_2$ walls (a) 200~nm and
(b) 2~$\mu$m.
}
\end{figure}

Next we consider the relative difference in the magnitudes of the Casimir pressure
on an addition of magnetite nanoparticles to kerosene as a function of
nanoparticle diameter.  The computational results are shown in Fig.~\ref{fg5}
by the solid and dashed lines computed with disregarded and included conductivity
of magnetite at low frequencies, respectively, for separation between the walls
(a) $200$~nm and (b) $2~\mu$m.

{}From Fig.~\ref{fg5}(a,b) it is seen that at both separations considered the
relative change in the magnitude of the Casimir pressure is a monotonously
increasing function of the nanoparticle diameter $d$ which changes its sign and
takes the zero value for some $d$.  Thus, from Fig. 5(a) one concludes that at
$a=200$ nm $\delta|P|=0$ for $d=16.6$ nm if the conductivity of magnetite at
low frequencies is disregarded in computations and for $d=16.8$ nm if it is
taken into account.  This means that at $a=200$ nm an inclusion in kerosene
of nanoparticles with some definite diameter does not make any impact on the
Casimir pressure.  According to Fig. 5(b), similar situation holds at $a=2~\mu$m.
Here, $\delta|P|=0$ for $d=12.9$ and $13.3$ nm depending on whether the conductivity
of magnetite at low frequencies is disregarded or included in computations.

The obtained results can be qualitatively  explained by the fact that for
magnetic materials with $\mu>1$ the magnitude of the Casimir pressure is always
larger, as compared to materials with $\mu=1$.  On the other hand, the presence
of magnetite nanoparticles in kerosene influences on its dielectric permittivity
in such a way that the magnitude of the Casimir pressure decreases.  These two
tendencies act in the opposite directions and may nullify an impact of magnetic
nanoparticles with some definite diameter on the Casimir pressure.

\section{The Case of Water-based Interlayer}

In this section we consider the Casimir pressure in the three-layer system where
an interlayer is formed by the water-based ferrofluid.  Water is of frequent use
as a carrier liquid (see, e.g., Refs. \cite{b57,b58,b59}).  The dielectric
permittivity of water along the imaginary frequency axis is well described in
the oscillator representation \cite{b60}
\begin{equation}\label{eq20}
  \varepsilon_c(i\xi)=1+\frac{B}{1+\xi\tau}+\sum_{j=1}^{11}
\frac{C_j}{1+(\frac{\xi}{\omega_j})^2+g_j\frac{\xi}{\omega_j^2}},
\end{equation}
\noindent
where the second term on the right hand-side of this equation describes the
contribution from the orientation of permanent dipoles.  Specifically, for water
one has $B=76.8$ and $1/\tau=1.08\times10^{11}$~rad/s.
The oscillator terms with $g=1,\,2,...,\,6$ represent the effects of
electronic polarization.  The respective oscillator frequencies belong
to the ultraviolet spectrum:
$\omega_j=1.25\times10^{16},\ 1.52\times10^{16},\ 1.73\times 10^{16},\
2.07\times 10^{16},\ 2.70\times 10^{16}$ and $3.83\times 10^{16}$~rad/s.
The oscillator strength and relaxation parameters of these oscillators are
given by: $C_j=0.0484,\ 0.0387,\ 0.0923,\ 0.344,\ 0.360,\ 0.0383$ and
$g_j=0.957\times 10^{15},\ 1.28\times 10^{15},\ 3.11\times 10^{15},\
5.92\times 10^{15},\ 11.1\times 10^{15},\ 8.11\times 10^{15}$ rad/s.
The terms of Eq.~(\ref{eq20}) with $j=7,\,8,...,\,11$ represent the effects
of ionic polarization and their frequencies belong to the infrared spectrum:
 $\omega_j=0.314\times 10^{14},\ 1.05\times 10^{14},\ 1.40\times 10^{14},\
3.06\times 10^{14},\ 6.46\times 10^{14}$~rad/s.  The respective oscillator
strengths and relaxation parameters take the following values:
 $C_j=1.46,\ 0.737,\ 0.152,\ 0.0136,\ 0.0751$ and
$g_j=2.29\times 10^{13},\ 5.78\times 10^{13},\ 4.22\times 10^{13},\
3.81\times 10^{13},\ 8.54\times 10^{13}$~rad/s \cite{b60}.

Using the dielectric permittivity $\varepsilon_c$ of water (\ref{eq20}) and
the dielectric permittivity of magnetite nanoparticles $\varepsilon_m$ given
by the top line in Fig.~\ref{fg1}, the permittivity $\varepsilon_{\rm ff}(i\xi)$
of the water-based ferrofluid with $\Phi=0.05$  fraction of nanoparticles is
obtained from Eq.~(\ref{eq11}).  It is shown by the line labeled 2 in
Fig.~\ref{fg2} as a function of the imaginary frequency normalized to the
first Matsubara frequency.  The dielectric permittivity of the water-based
ferrofluid at zero Matsubara frequency is equal to $77.89$ if the conductivity
of magnetite at low frequencies is disregarded and to $93.97$ if it is taken
into account in calculations.

The magnitude of the Casimir pressure between two SiO$_2$ walls through the
water-based ferrofluid was computed
similar to Sec.~III by using Eqs.~(\ref{eq17}) and
(\ref{eq18}) and all respective dielectric permittivities and magnetic
permeabilities defined along the imaginary frequency axis.
The computational results are shown in Fig.~\ref{fg3} as  functions of separation
by the pair of solid lines labeled 2 where the lower and upper lines are computed
for magnetite nanoparticles with $d=10$ and $20$~nm diameter, respectively.
Similar to Sec.~III, the conductivity of magnetite at low frequencies was
first disregarded.  Then the computations have been repeated with taken
into account conductivity.  The obtained results are shown by the pair of
dashed lines labeled 2.  As is seen in Fig.~\ref{fg3}, an account of the conductivity
of magnetite makes only a minor impact on the Casimir pressure.
Thus, for $d=10$ nm, $a=200~$nm one obtains $P=-19.82$~mPa and
$\tilde{P}=-20.61$~mPa with disregarded and included conductivity of magnetite,
respectively.  At larger separation $a=2~\mu$m similar results are $P=-1.964\times10^{-2}$~mPa and $\tilde{P}=-2.043\times10^{-2}$~mPa.
For nanoparticles of $d=20$~nm diameter we find $P=-24.62$~mPa and
$\tilde{P}=-25.42$~mPa at $a=200$~nm and $P=-2.444\times10^{-2}$~mPa, $\tilde{P}=-2.524\times10^{-2}$~mPa at $a=2~\mu$m.

\begin{figure}[b]
\vspace*{-4cm}
\centerline{\hspace*{2.5cm}
\includegraphics{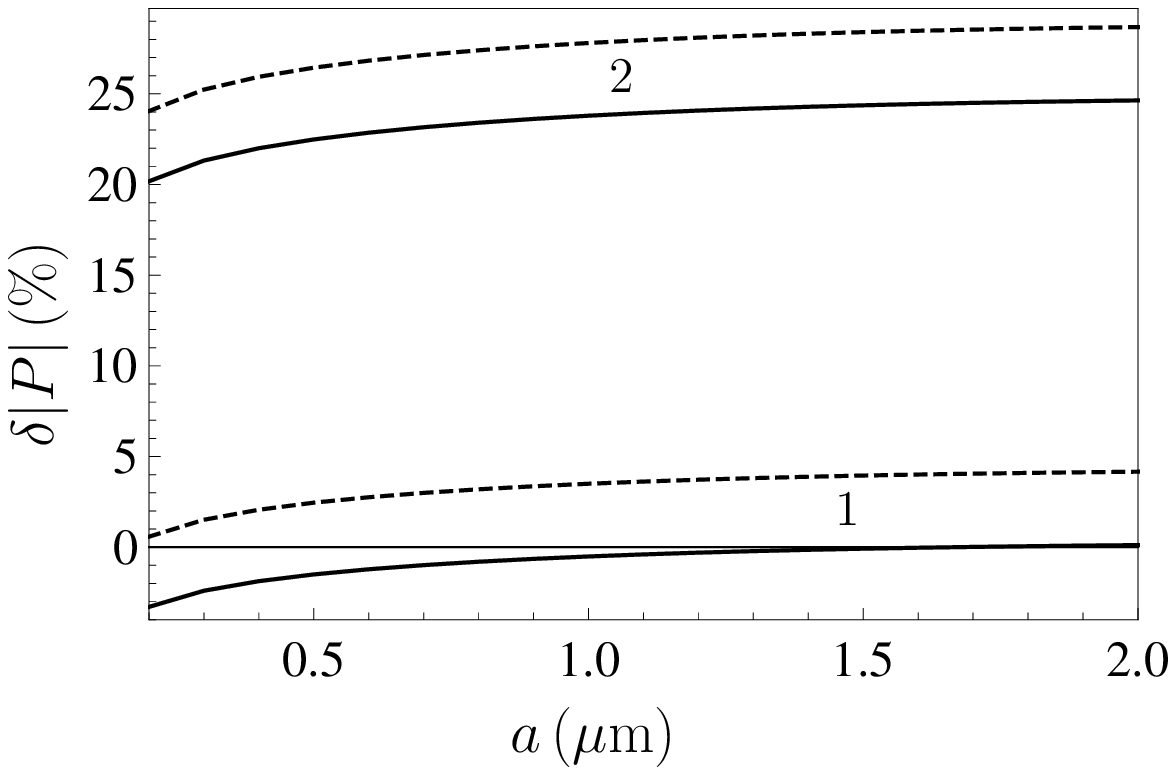}
}
\vspace*{-9.5cm}
\caption{\label{fg6}
The relative change in the magnitude of the Casimir pressure on addition of the
5$\%$ fraction of magnetite nanoparticles to water is shown as the function
of separation by the pairs of solid and dashed lines labeled 1 and 2 for
nanoparticles with $d=10$ and 20~nm diameter, respectively.  In each pair, the
solid and dashed lines are computed with disregarded and included conductivity
of magnetite at low frequencies, respectively.
}
\end{figure}

The relative impact of magnetite nanoparticles on the Casimir pressure in
three-layer system with a water interlayer can be calculated by Eq.~(\ref{eq19})
where $|P_{\rm ker}|$ should be replaced with $|P_{\rm wat}|$ found for a pure
water stratum sandwiched between two SiO$_2$ walls.  The computational results
are shown in Fig.~\ref{fg6} as functions of separation by the lines labeled 1
and 2 for nanoparticles with $d=10$ and $20$~nm diameter, respectively.
As above, the solid and dashed lines in each pair are computed with disregarded
and included conductivity of magnetite, respectively.

{}From Fig.~\ref{fg6} it is seen that on addition of nanoparticles with
$d=20$~nm diameter to water the magnitude of the Casimir pressure increases
independently of whether the conductivity of magnetite is disregarded or included
in computations.  The same is true for nanoparticles of $d=10$~nm diameter, but
only under a condition that the conductivity of magnetite  is taken into account.
If the conductivity is disregarded for nanoparticles with $d=10$~nm (the solid
line labeled 1),  the quantity $\delta|P|$ takes the negative values over the
wide separation range, i.e., on addition of magnetite nanoparticles to water
the magnitude of the Casimir pressure becomes smaller.

At $a=200$~nm one obtains $\delta|P|=-3.3\%$ and $20.4\%$ for nanoparticles with
$d=10$ and 20~nm diameter, respectively, and $\delta|\tilde{P}|=0.58\%$
and $24.3\%$  for the same respective diameters.  At $a=2 ~\mu$m similar results
are the following: $\delta|P|=0.12\%$ and $24.9\%$, and $\delta|\tilde{P}|=4.2\%$
and $29.0\%$ for $d=10$ and 20~nm, respectively.

\begin{figure}[b]
\vspace*{-1cm}
\centerline{\hspace*{2.5cm}
\includegraphics{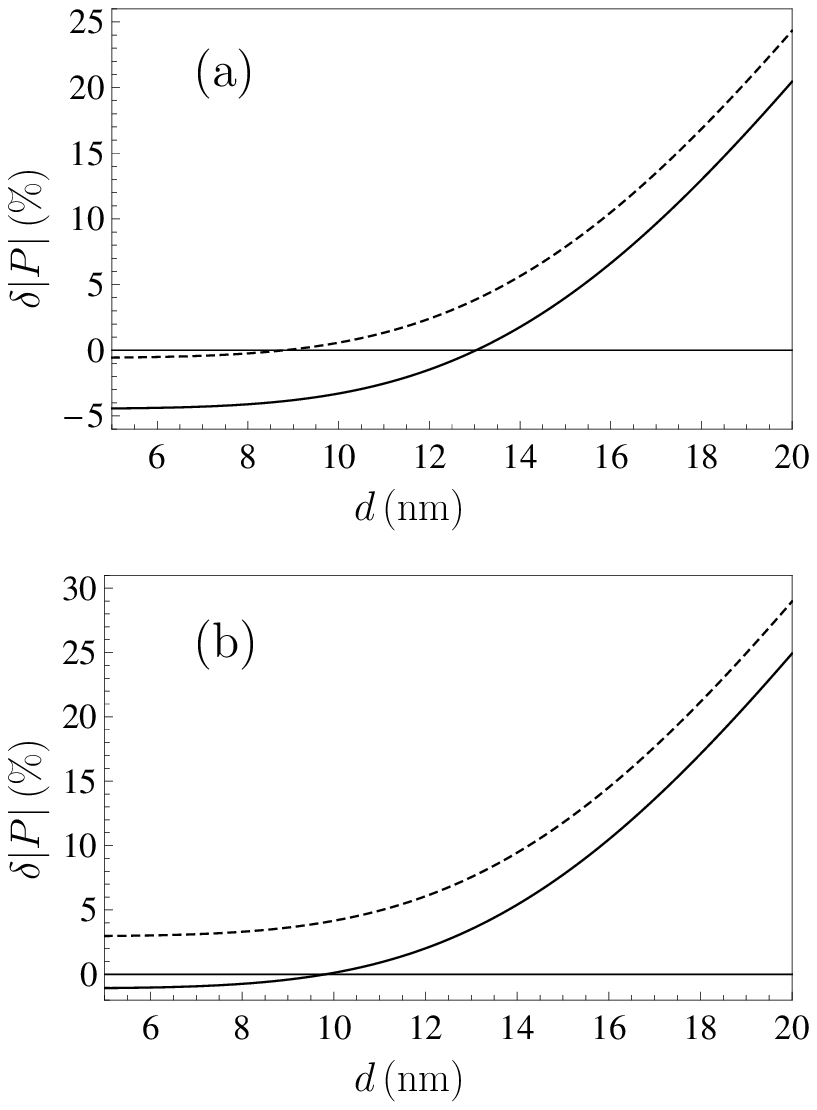}
}
\vspace*{-13.5cm}
\caption{\label{fg7}
  The relative change in the magnitude of the Casimir pressure on addition of the
  5$\%$ fraction of magnetite nanoparticles to water is shown as the function
  of nanoparticle diameter by the solid and dashed lines computed with disregarded
  and included conductivity of magnetite at low frequencies, respectively, for
  separation between SiO$_2$ walls (a) 200~nm and (b) 2~$\mu$m.
}
\end{figure}

In the end of this section, we calculate the relative difference in Casimir
pressures $\delta|P|$ for a water-based ferrofluid as a function of nanoparticle
diameter.  In Fig.~\ref{fg7} the computational results are presented by the solid
and dashed lines computed with disregarded and included conductivity of
magnetite, respectively, for separation between the walls $a=200$~nm in
Fig.~\ref{fg7}(a) and $a=2~\mu$m in Fig.~\ref{fg7}(b).

As can be seen in Fig.~\ref{fg7}, for a water-based ferrofluid the quantity
$\delta|P|$ is again the monotonously increasing function of nanoparticle diameter
$d$ which changes its sign for some value of $d$.  According to Fig.~\ref{fg7}(a),
at $a=200$~nm and Casimir pressure does not change on an addition of
magnetite nanoparticles with $d=13$~nm diameter to water
if the conductivity at low frequencies
is disregarded.  If it is included in computations, the value of this diameter is
reduced to $d=8.8$~nm.  If the separation distance between SiO$_2$ walls is
$a=2~\mu$m, an addition to water of magnetite nanoparticles with $d=9.8$~nm diameter
does not make an impact to the Casimir pressure under a condition that the
conductivity of magnetite is disregarded.  If this conductivity is included
in computations, an addition of nanoparticles of any diameter to water influences
the Casimir pressure in the three-layer system.  The qualitative explanation of
this effect is the same as considered in Sec.~III for the kerosene-based ferrofluids.

\section{Conclusions and Discussion}

In the foregoing, we have considered the Casimir pressure in three-layer systems
where the intervening stratum processes magnetic properties.  This subject has
assumed importance in the context of ferrofluids and their extensive use in
micromechanical sensors and other  prospective applications discussed in
Sec.~I.  Taking into account that the magnetic properties of ferrofluids and
determined by some fraction of magnetic nanoparticles added to a nonmagnetic
(carrier) liquid, we have investigated an impact of such nanoparticles with
different diameters of the Casimir pressure.

After presenting the general formalism of the Lifshitz theory adapted to this case,
we have found the dielectric permittivity and the magnetic permeability of
 magnetite nanoparticles along the imaginary frequency axis on the basis of
 available optical data.  Specific computations have been performed for the
 kerosene- and water-based ferrofluids which are most commonly studied in the
 literature.  For this purpose, we have constructed the dielectric permittivity
of kerosene using its measured optical properties in the microwave and infrared
regions and employed the familiar representation for the dielectric permittivity
of water.  These permittivities have been combined with the permittivity of
magnetite nanoparticles by using the Rayleigh mixing formula to obtain the
dielectric permittivities of ferrofluids with 5$\%$ concentration of nanoparticles.

The Casimir pressure was computed for the three-layer systems consisting of two
 parallel SiO$_2$ walls with an intervening stratum of either kerosene- or
 water-based ferrofluid as a function of separation between the walls.  We have
 also computed the relative difference in the magnitudes of the Casimir pressure
 on addition of the $5 \%$ fraction of magnetic nanoparticles to the pure kerosene
 and water as a function of separation and nanoparticle diameter.  It was shown that
this relative difference is rather large and should be taking into account.
As an example, for kerosene- and water-based ferrofluids with nanoparticles of
20~nm diameter sandwiched between two SiO$_2$ walls 2~$\mu$m apart,
the relative change in the magnitude of the Casimir pressure exceeds 140$\%$ and
25$\%$, respectively.

All computations have been performed in the framework of two theoretical approaches
to the Casimir force developed in the literature during the last twenty years and
used for comparison between experiment and theory.  It turned out that both of
these approaches lead to fairly close predictions for the magnitudes of the
Casimir pressure in three-layer systems through a ferrofluid interlayer.
In doing so, theoretical predictions for the relative change in the magnitude
of the Casimir pressure on addition of magnetic nanoparticles to a carrier liquid
is more sensitive to the approach used and may vary in the limits of several percent.

An interesting effect found for the three-layer systems with a ferrofluid
intervening stratum is that at fixed separation between the walls an addition of
magnetite nanoparticles with some definite diameter to a carrier liquid makes no
impact on the Casimir pressure between SiO$_2$ walls.  The respective diameters
are found for both kerosene- and water-based ferrofluids.  The quantitative
physical explanation for this effect is provided.

In conclusion, it may be said that the above results open  opportunities for
precise control of the Casimir force in the three-layer systems with  a magnetic
intervening stratum, which may be used in the next generation of
ferrofluid-based microdevices.

\section{Acknowledgments}

The work of V.~M.~M.~was partially supported by the Russian Government Program
of Competitive Growth of Kazan Federal University.

\newpage

\end{document}